\documentclass[fleqn,usenatbib]{mnras}

\usepackage[T1]{fontenc}

\DeclareRobustCommand{\VAN}[3]{#2}
\let\VANthebibliography\thebibliography
\def\thebibliography{\DeclareRobustCommand{\VAN}[3]{##3}\VANthebibliography}

\usepackage{graphicx}	% Including figure files
\usepackage{amsmath}	% Advanced maths commands
\usepackage{amssymb}	% Extra maths symbols

\title[Magnetotail Turbulence]{Observation of Inertial-range Energy Cascade within a Reconnection Jet in Earth's Magnetotail}

\author[R. Bandyopadhyay et al.]{
Riddhi Bandyopadhyay,$^{1}$\thanks{E-mail: riddhib@princeton.edu}\thanks{The majority of this work was completed at the Department of Physics and Astronomy, University of Delaware, Newark, DE 19716, USA.}
Alexandros Chasapis,$^{2}$
D. J. Gershman,$^{3}$
B. L. Giles,$^{3}$
\newauthor
C. T. Russell,$^{4}$ 
R. J. Strangeway,$^{4}$
O. Le Contel,$^{5}$
M. R. Argall$^{6}$
and J. L. Burch$^{7}$
\\
$^{1}$Department of Astrophysical Sciences, Princeton University, Princeton, NJ 08544, USA\\
$^{2}$Laboratory for Atmospheric and Space Physics, University of Colorado, Boulder, CO 80303, USA\\
$^{3}$NASA Goddard Space Flight Center, Greenbelt, MD 20771, USA\\
$^{4}$University of California, Los Angeles, CA 90095-1567, USA\\
$^{5}$Laboratoire de Physique des Plasmas, CNRS/Ecole Polytechnique/Sorbonne Universit\'e/Universit\'e Paris-Saclay/Observatoire de Paris, Paris, France\\
$^{6}$University of New Hampshire, Durham, NH 03824,USA\\
$^{7}$Southwest Research Institute, San Antonio, TX 78238-5166, USA}

\date{Accepted XXX. Received YYY; in original form ZZZ}

\pubyear{2020}

\begin{document}
\label{firstpage}
\pagerange{\pageref{firstpage}--\pageref{lastpage}}
\maketitle

\begin{abstract}
 Earth's magnetotail region provides a unique environment to study plasma turbulence. We investigate the turbulence developed in an exhaust produced by magnetic reconnection at the terrestrial magnetotail region. Magnetic and velocity spectra show broad-band fluctuations corresponding to the inertial range, with Kolmorogov $-5/3$ scaling, indicative of a well developed turbulent cascade. We examine the mixed, third-order structure functions, and obtain a linear scaling in the inertial range. This linear scaling of the third-order structure functions implies a scale-invariant cascade of energy through the inertial range. A Politano-Pouquet third-order analysis gives an estimate of the incompressive energy transfer rate of $\sim 10^{7}~\mathrm{J\,kg^{-1}\,s^{-1}}$. This is four orders of magnitude higher than the values typically measured in 1 AU solar wind, suggesting that the turbulence cascade plays an important role as a pathway of energy dissipation  during reconnection events in the tail region. 
\end{abstract}

\begin{keywords}
turbulence -- magnetohydrodynamics (MHD) -- plasmas -- reconnection
\end{keywords}

%%%%%%%%%%%%%%%%%%%%%%%%%%%%%%%%%%%%%%%%%%%%%%%%%%

%%%%%%%%%%%%%%%%% BODY OF PAPER %%%%%%%%%%%%%%%%%%

\section{Introduction}
Magnetic reconnection and turbulence are two fundamental processes observed in many natural systems~\citep{Vasyliunas1975JGR, Parker1979book, Taylor86}. Magnetic reconnection is the process by which two magnetic field lines reconfigure their topology, converting magnetic energy into kinetic and thermal energy~\citep{Parker1957JGR, SonnerupEA84, SchindlerEA88}. Turbulence is generally characterized by the presence of broadband fluctuations in space and time. The non-linear interaction between the eddies in a turbulent system, leads to transfer of energy from large to small scales, ultimately dissipating into heat - a process known as \textit{energy cascade}~\citep{Richardson1922book,Karman1938PRSL,Kolmogorov1941a,Frisch1995Book}. {Using two-dimensional simulations, magnetic reconnection is shown to occur at small-scales in thin current sheets, as a part of the \emph{turbulent cascade} process, dissipating magnetic energy into flow and thermal energy~\citep{Servidio2009PRL}}. On the other hand, large-scale reconnection, such as the ones observed in the Earth's magnetopause and magnetotail, can \emph{drive} the plasma turbulence, injecting large-scale energy into the system. The interplay of the two processes have been a subject of increasing interest in space physics research~\citep{Matthaeus1986PoF,EastwoodEA09,Matthaeus2011SSR,Chasapis2017ApJ,Shay2018PoP}.

Magnetic reconnection in the  magnetotail region is understood as one of the most important mechanisms to provide energy for auroral substorms. {The magnetotail plasma is known to exhibit characteristics of turbulent fluctuations~\citep{Borovsky1997JPP}. While turbulence in the solar wind has been extensively studied with the
help of a large fleet of spacecraft~\citep[see reviews by][]{Bruno2013LRSP, Verscharen2019LRSP}, the literature focusing on magnetotail turbulence remains sparse~\citep{Voros2004JGR, Stepanova2009AG, Zimbardo2010SSR}.} Nevertheless, the  properties of turbulence, developed in the outflows of reconnection, particularly in the context of magnetotail reconnection, have become a topic of recurrent interest in the recent years~\citep{Huang2012GRL,Osman2015ApJL,Pucci2017ApJ}. 

\begin{figure*}
	\begin{center}
		\includegraphics[width=\linewidth]{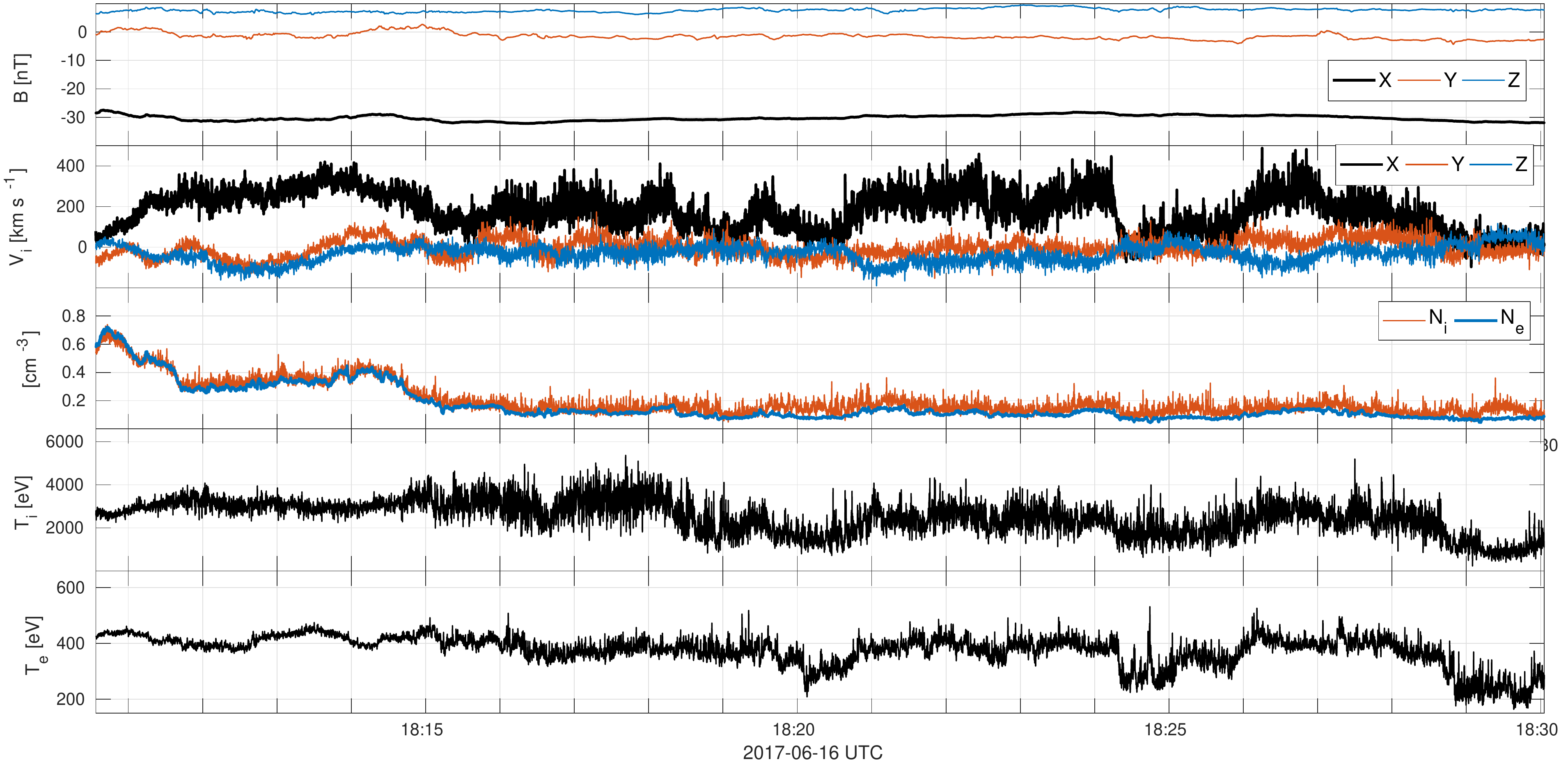}
		\caption{Overview of the MMS observations selected for this study. The data shown are from the FGM and FPI instruments on-board the MMS1 spacecraft. The panels plot the magnetic field components in GSE coordinates, the ion velocity, the ion and electron density,   the ion temperature, and the electron temperature.}
		\label{fig:overview}
	\end{center}
\end{figure*}

A main characteristic of a well-developed turbulent system is the presence of a scale-invariant energy cascade at the inertial range. This constant energy flux in the inertial scales is the basis of Kolmogorov's power-law spectrum~\citep{Kolmogorov1941a}. {But a third-order statistics, first developed in hydrodynamics~\citep{Kolmogorov1941c,MoninYaglom-vol1,MoninYaglom-vol2,Frisch1995Book} and later extended to magnetohydrodynamic (MHD) turbulence by \cite{Politano1998GRL,Politano1998PRE}, reveal the existence and value of the  inertial-scale energy flux more directly.} The Politano-Pouquet law, in its isotropic form, has been applied to estimate the turbulent heating rate 
in the solar-wind~\citep[e.g.,][]{MacBride2008ApJ, Sorriso-Valvo2007PRL}, magnetosheath~\citep{Bandyopadhyay2018bApJ, Hadid2018PRL}, and magnetospheric boundary layer~\citep{Sorriso-Valvo2019PRL}. However, due to instrumental limitations of past missions, direct observations of inertial-range energy transfer in the magnetotail region have not been possible. In this paper, we investigate the MHD-scale energy transfer rate in a reconnection exhaust detected by the Magnetospheric Multiscale (MMS) mission~\citep{Burch2016SSR} in Earth's magnetotail region.

\section{MMS Observations} \label{sec:overview}
\begin{table}
	\caption{Description of some plasma parameters. Data obtained on 2017 June 16 from 18:10:03 to 18:29:33 UT. The magnetic-field rms fluctuation amplitude is calculated as $B_{\mathrm{rms}} = \sqrt{\langle |\mathbf{B}(t) - \langle \mathbf{B} \rangle|^2 \rangle}$. Ion inertial length $d_{\mathrm{i}}$, electron inertial length $d_{\mathrm{e}}$, ion velocity $\mathbf{V}$, and the proton plasma beta $\beta_{\mathrm{p}} = v_{\mathrm{th}}^2/V_{\mathrm{A}}^2$ are also reported.} 
	\label{tab:overview}
	%\centering
	\begin{center}
		\begin{tabular}{l c c c c c c r}
			\hline
			$|\langle \mathbf{B} \rangle|$ & $|\langle \mathbf{V}_{\mathrm{A}} \rangle|$ & $\frac{B_{\mathrm{rms}}} {|\langle \mathbf{B} \rangle|}$ 
			& $\langle n_{\mathrm{i}} \rangle$ 
			& $d_{\mathrm{i}}$ 
			&$|\langle \mathbf{V} \rangle|$ & $V_{\mathrm{rms}}$
			 & $\beta_{\mathrm{p}}$ \\
			{($\rm nT$)} & {($\rm km/s$)} &  &  {(${\rm cm^{-3}}$)} & {($\rm km$)} & {($\rm km/s$)} & {($\rm km/s$)} &  \\
			\hline
			31 & 1554 & 0.05 & 0.22 & 466 & 171 & 100 & 0.24\\
			\hline
		\end{tabular}
	\end{center}
\end{table}

MMS is designed to study magnetic reconnection, turbulence, and particle acceleration in Earth's magnetosphere at high spatial and temporal
resolution with four satellites, in a tetrahedral formation. We use burst resolution MMS data obtained in the magnetotail region on 2017 June 16 from 18:10:03 to 18:29:33 UTC. The separation of the four MMS spacecraft was $\approx 35$ km, which is much smaller than the ion inertial length ($d_\mathrm{i} \approx 466$ km). 
{We use magnetic field  measurements from the Fluxgate~\citep[FGM;][]{Russell2016SSR} and search-coil~\citep[SCM;][]{LeContel2016SSR}
magnetometers and particle moments from the Fast Plasma
Investigation~\citep[FPI;][]{Pollock2016SSR} instrument.}
An overview of the interval is shown in Fig.~\ref{fig:overview}. Corrections have been applied to the FPI/DIS moments according to the method described in ~\cite{Gershman2019JGR} to account for the penetrating radiation effect. From the top panel of Fig.~\ref{fig:overview},  a very strong mean-magnetic field is present in the $-X_{\mathrm{GSE}}$ direction. The value of $B_{\mathrm{rms}} / |\langle \mathbf{B} \rangle|$ is 0.05, which indicates that the mean-magnetic field is much stronger than the turbulent fluctuations. This strong mean field is expected to generate spectral anisotropy~\citep{Shebalin1983JPP,Oughton1994JFM}. This strong $B_{\rm X}$ field indicates that MMS is mostly located at the edge of the plasma sheet where the beta is rather small whereas near the equator $B_{\rm total}$ is smaller and the plasma beta can be high up to 50-100. So, the observations are characteristic of the edge of the plasma sheet.

A major challenge in performing the usual turbulence studies in the magnetotail region is that, unlike the solar wind or magnetosheath, often there is no clear bulk flow in these plasmas. Therefore, in such a case, the usual interpretation of spacecraft time series data as a one-dimensional spatial sample, by Taylor's frozen-in hypothesis~\cite{Taylor1938PRSLA}, is not applicable. This is one of the primary reasons for the relative scarcity of such observations in magnetotail turbulence ~\citep{Borovsky1997JPP, Borovsky2003JGR, Ergun2018GRL}, compared to solar wind turbulence~\citep{Bruno2013LRSP}. Nevertheless, in the current interval, due to the strong reconnection exhaust, there is a relatively strong and steady flow in the $X_{\mathrm{GSE}}$ direction. 

{The mean proton temperature is $T_{\mathrm{i}} \sim 3000$ eV, which is typical for magnetotail plasma ($\sim$ 1-10 keV), but it is about 200 times higher than the usual ion temperature in the solar wind, and about 30 times higher than the typical ion temperature in the magnetosheath.} This high temperature indicates that a strong heating process may be going on in the plasmas, possibly due to reconnection heating and turbulent cascade. The average density is about $0.22$ cm$^{-3}$, resulting in a plasma beta of $\beta_{\mathrm{p}} = 0.24$, smaller than unity.

\subsection{Taylor Hypothesis} \label{sec:taylor}
\begin{figure}
	\begin{center}
		\includegraphics[scale=1.0]{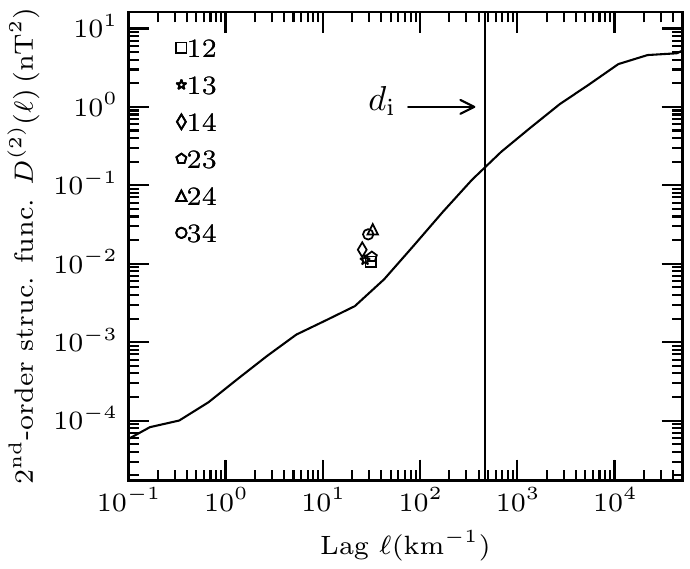}
		\caption{Test of Taylor's Hypothesis using second-order structure function of the magnetic field. The solid line is obtained using the Taylor's hypothesis, and averaged over the 4 MMS spacecraft. The symbols are the values computed using a two-spacecraft method. The solid, vertical  line represents the Taylor-shifted ion-inertial length scale $d_{\mathrm{i}}$.}
		\label{fig:taylor}
	\end{center}
\end{figure}
From single spacecraft measurements, one obtains a time series sample. If the average flow speed of the plasma with respect to the spacecraft is sufficiently faster than the Alfv\'en speed, one can assume Taylor hypothesis~\citep{Taylor1938PRSLA} to convert the temporal ($\tau$) measurements to spatial ($\ell$) measurement : $\ell = V_{\mathrm{f}} \tau$, where $V_{\mathrm{f}}$ is the flow speed of the plasma. {For the chosen MMS interval, however, the Alfv\'en speed, calculated from the mean magnetic field, is 9 times larger than the mean flow speed (table~\ref{tab:overview}) and $V_{\mathrm{rms}}/V_{\mathrm{f}}=0.58$}. Therefore, it is not obvious that the Taylor hypothesis would be applicable here. To test the validity of the Taylor hypothesis, we take advantage of a multi-spacecraft method. We calculate the second-order structure function of the magnetic field,
\begin{eqnarray}
D^{(2)}(\ell) = \langle |\mathbf{B}(\mathbf{x}+\mathbf{\ell}) - \mathbf{B}(\mathbf{x})|^2 \rangle, \label{eq:s2}
\end{eqnarray}
using the spatial lags between six 2-spacecraft pairs, and compare them with the same calculated from the time series, assuming Taylor's frozen-in flow with a mean speed of 171 km/s. A FGM-SCM merged  (FSM) data product with a crossover between the instruments from 4 to 7 Hz is used~\citep[][in preparation]{Argall2018ArXiv} for this purpose.

Figure~\ref{fig:taylor} plots the second-order structure function using the two methods. {Despite having a considerably larger Alfv\'en speed compared to the mean flow speed}, Taylor's frozen-in theorem appears to roughly hold in the sub-proton scales. {We recall that $V_{\mathrm{f}} >> V_{\mathrm{A}}$, is the most stringent condition for the case of fluctuations which are propagating Alfv\'en wave. If the fluctuations represent well-developed turbulence, this condition is somewhat relaxed~\citep{Matthaeus1982aJGR}. Eddy-turnover times for turbulent magnetic structures of size $\lambda \sim 1/k$ may be estimated~\citep{Batchelor1953book, Dobrowolny1980PRL} as $t_{\lambda} \sim 1/(b_{k} k) $, where $b_k$, is the Fourier-series amplitude of magnetic-field, measured in Alfven units, at wave number $k$. The fluctuations are frozen in if the characteristic time, $t_{\lambda}$, for dynamical evolution of structure of size $\lambda$ is much greater than the convection time of the structure with speed $V_{\mathrm{f}}$: $t \sim \lambda / V_{\mathrm{f}}$. Therefore, the condition $t_{\lambda} \gg t$ will be satisfied if
$V_{\mathrm{f}} \gg b_k$.
The magnitude of $b_k$ will depend on $k$ but, will be smaller than the rms fluctuation $\delta V_{A} = 78$km/s. Given that  $\delta V_{A} / V_f = 0.5$, one may expect that the 'frozen-in' approximation holds for all MHD-scale fluctuations.}
\begin{figure}
	\begin{center}
		\includegraphics[scale=0.9]{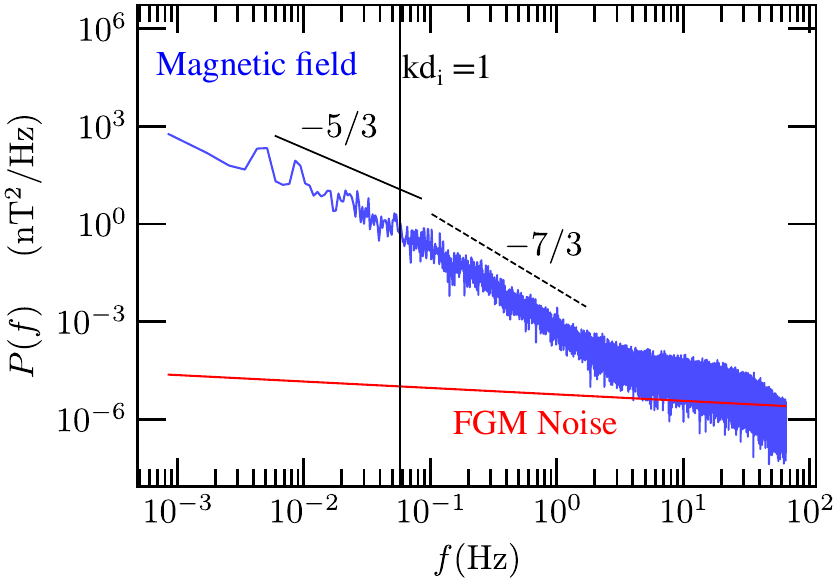}
		\includegraphics[scale=0.9]{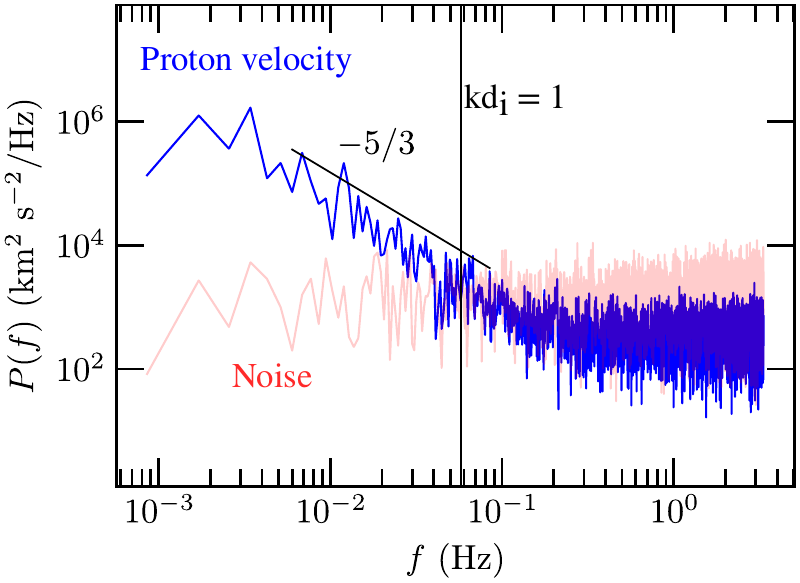}
		\caption{Spectra of magnetic field (top) and ion velocity field (bottom). Kolmogorov scaling $\sim f^{-5/3}$ is shown for reference. The solid vertical line represents $k d_i = 1$ with the wave vector $k \simeq (2\pi f)/|\langle \mathbf{V} \rangle|$.}
		\label{fig:spec_b}
	\end{center}
\end{figure}
{Similar observations were found in some other sub-Alfv\'enic flows~\citep{Stawarz2016JGR, Sorriso-Valvo2019PRL}. Fig.~\ref{fig:taylor} suggests that the multi-spacecraft measurements are about $\sim 10$ times higher than the Taylor-based ones. However, we expect the quality of the frozen-in approximation to improve at the inertial scales, compared to kinetic scales. Since inertial-range fluctuations are of interest here, we proceed by assuming that the Taylor hypothesis is reasonably valid with a flow speed of $171$ km/s.}

\subsection{Spectra} \label{sec:spectra}
To study the nature of MHD-scale turbulent fluctuations in the selected interval, we begin by computing the power spectra for magnetic and proton-velocity field. Note that although we used merged FSM magnetic-field data to test the validity of Taylor hypothesis in section~\ref{sec:taylor}, to investigate the nature of MHD-scale turbulence, the FGM data resolution suffices. Figure~\ref{fig:spec_b} shows that magnetic-field spectrum computed from FGM data and proton velocity spectrum computed from FPI data. Both spectra exhibit a clear inertial-range with Kolmogorov $-5/3$ scaling. The magnetic-field spectrum shows a steepening to $\sim -7/3$ slope near the ion-inertial scale $d_{\mathrm{i}}$, and then the spectrum flattens due to noise $\sim 5$ Hz. The velocity spectrum becomes dominated by the counting-statistics noise~\citep{Gershman2018PoP}  near $k d_{\mathrm{i}} = 1$, so the steepening can not be  observed in this case. The broad inertial-range scaling for the magnetic and velocity field indicate a well-developed, classical Kolmogorov-type turbulence with a scale-invariant flux in the inertial range~\citep{Frisch1995Book,Biskamp2003Book,Hadid2015ApJL,Huang2014ApJL}.
	
\section{Politano-Pouquet Energy Cascade} \label{sec:eps2}
To estimate the energy transfer rate ($\epsilon$) in the inertial scale, we use the Kolmorogov-Yaglom law, extended to isotropic MHD~\citep{Politano1998GRL, Politano1998PRE},
\begin{eqnarray}
Y^{\pm}(\ell) = - \frac{4}{3} \epsilon^{\pm} \ell \label{eq:3ord},
\end{eqnarray}
where $Y^{\pm}(\ell)=\langle \mathbf{\hat{\ell}}\cdot \Delta \mathbf{Z}^{\mp}(\mathbf{x, \ell}) |\Delta \mathbf{Z}^{\pm}(\mathbf{x, \ell})|^2 \rangle$, are the mixed third-order structure functions. The 
Elsasser variables are defined as $\mathbf{Z}^{\pm} = \mathbf{V} \pm {\mathbf{B}}/{\sqrt{\mu_{0}\,m_{\mathrm{p}}\,n_{\mathrm{i}}}}$. $\Delta \mathbf{Z}^{\pm}(\mathbf{x, \ell})$ are the increments of the Elsasser variables at position $\mathbf{x}$, for lag $\mathbf{\ell}$: $\Delta \mathbf{Z}^{\pm}(\mathbf{x, \ell}) = \mathbf{Z}^{\pm}(\mathbf{x + \ell}) - \mathbf{Z}^{\pm}(\mathbf{x})$. Here, $\mu_0$ is the magnetic permeability of vacuum, $m_{\mathrm{p}}$ 
is the proton mass, respectively and $n_{\mathrm{i}}$ is the proton number density. $\epsilon^{\pm} = - \mathrm{d} (Z^{\pm})^2/\mathrm{d}t$ are the cascade rate for each Elsasser variable. The energy cascade rate is obtained by averaging the two: $\epsilon = - \mathrm{d} E/\mathrm{d}t = (\epsilon^{+} + \epsilon^{-})/2$ {, where $E$ is the total (magnetic+kinetic) fluctuation energy: $E=v^2+b^2$.}

Equation~(\ref{eq:3ord}), known as the third-order law or the Politano-Pouquet law, has been the standard approach in estimating the turbulent cascade rate in space plasmas. We note here that second-order statistics has also been used to evaluate the energy flux, although with some additional assumptions and parameters~\citep{Verma1995JGR,Verma1996JGR}. Further, compressibility has been taken into account in some variations of the Yaglom law~\citep[e.g.,][]{Banerjee2013PRE}. Nevertheless, here we use the classical, incompressible Politano-Pouquet law to estimate the incompressive energy cascade. The third-order law, in its isotropic form as in equation~\ref{eq:3ord}, is derived assuming isotropy. However, even in strongly anisotropic systems, (as in \cite{Stawarz2009ApJ} and \cite{Osman2011PRL}), the results have been quite comparable with the isotropic case.

\begin{figure}
	\begin{center}
		\includegraphics[scale=1.0]{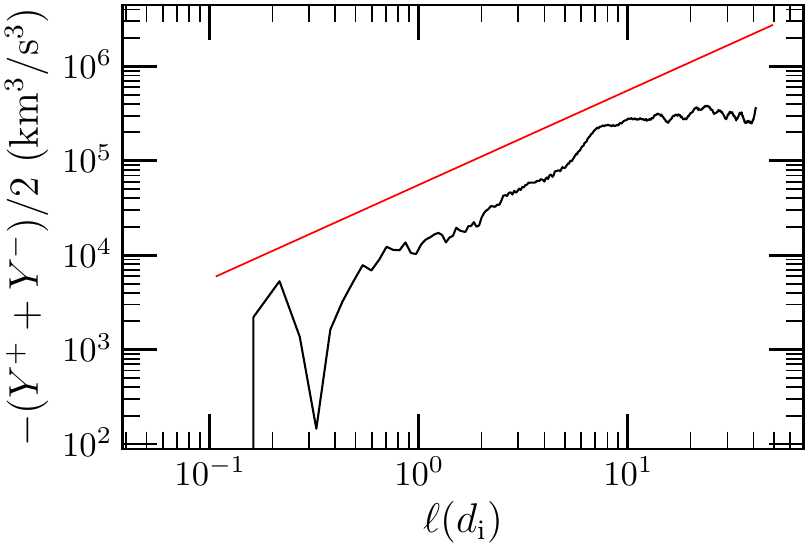}
		\caption{The average of the two mixed third-order structure functions. A linear scaling is shown for reference.}
		\label{fig:ypm}
	\end{center}
\end{figure}

For adequate statistics, we average the data from all four MMS spacecraft, and then calculate the structure functions. Figure~\ref{fig:ypm} 
 plots the average mixed third-order structure functions as a function of Taylor-shifted lag. {An approximately linear scaling} is indeed observed in the range $\sim 1-5\, d_{\mathrm{i}}$, and this approximate linear scaling is used to estimate an inertial-range energy flux. 
{The range of approximate linear scaling is rather short, spanning less than a decade. This short range is indicative of narrow scale separation and small Reynolds number of this system. Nevertheless, fitting a straight line in the inertial range we obtain,}
\begin{eqnarray}
\epsilon = (24.2 \pm 0.4) \times 10^{6} \mathrm{~J\,kg^{-1}\,s^{-1}} \label{eq:eps2}.
\end{eqnarray}

The uncertainty is obtained by fitting a straight-line of unit slope within the range $\ell=1$ to $5\, d_{\mathrm{i}}$. {For comparison, we note that the corresponding value in the 1 AU pristine solar wind~\footnote{Pristine solar wind, here, refers to the solar wind outside the bow shock, free from any shock contamination, such as reflected ions or upstream waves.} is $\sim 10^3\,\mathrm{~J\,kg^{-1}\,s^{-1}}$~\citep{Sorriso-Valvo2007PRL},} and that in the magnetosheath plasma is $\sim 10^6\,\mathrm{~J\,kg^{-1}\,s^{-1}}$~\citep{Bandyopadhyay2018bApJ, Hadid2018PRL}.

%\clearpage

\section{Summary and Discussion}
We present an estimate of turbulent energy cascade rate in a reconnection jet detected by MMS spacecraft. This is the first report of a Politano-Pouquet scaling, and consequently, of an energy decay estimate in a magnetotail reconnection exhaust. Previous studies had already established intermittency and energy conversion between particles and electromagnetic fields in similar systems~\citep{Huang2012GRL, Osman2015ApJL}. This study provides a next step in those direction. The analysis is facilitated by a fast, steady flow due to the reconnection jet, which allows us to use the Taylor hypothesis. {The possible validity of Taylor-hypothesis is argued using the two-spacecraft measurements.} Prior to MMS, such cross-checking was not possible, and studies had to rely on indirect ways of cross checking such as {the proximity of between the spectral-break and proton kinetic scales (gyro radius or ion-inertial length) to validate the Taylor hypothesis. This is based on earlier studies~\citep[e.g.,][]{ChenEA14} that showed the ion-scale spectral break occurs at the larger of the ion gyroradius and ion-inertial length.} The very high energy transfer rate, evaluated using the third-order law, may be responsible for the temperature profile in the magnetotail region, as has been suggested for other systems~\citep{MacBride2008ApJ, Saur2004ApJL}. The quantitative estimation of the heating rate, evaluated in this paper may provide key information for modeling of the Earth's  magnetosphere, and further our understanding of role of turbulence in magnetic reconnection. The measurements are recorded at the edge of the plasma sheet where beta is smaller than 1 and plasma velocity is often mostly parallel to $\mathbf{B}_0$. It remains to be seen how the conclusions will change with data gathered near the equatorial region where beta is much larger (at least $>1$) and plasma velocity mostly perpendicular to $\mathbf{B}_0$. Such an extension would represent a significant subsequent study.

\section*{Acknowledgments}
This research is partially supported by the MMS Mission
through NASA grant NNX14AC39G at the University of
Delaware. The French LPP involvement for the SCM instrument is supported by CNES and CNRS. We are grateful to the MMS instrument teams, especially SDC, FPI, and FIELDS, for cooperation and collaboration in preparing the data. 

\section*{Data Availability}
All MMS data are available at \url{https://lasp.colorado.edu/mms/sdc/}.

%%%%%%%%%%%%%%%%%%%%%%%%%%%%%%%%%%%%%%%%%%%%%%%%%%



\bibliographystyle{mnras}
%\bibliography{refs_riddhi,AG,HL,MP,QZ,refs_WHM} % if your bibtex file is called refs_riddhi.bib

\begin{thebibliography}{}
	\makeatletter
	\relax
	\def\mn@urlcharsother{\let\do\@makeother \do\$\do\&\do\#\do\^\do\_\do\%\do\~}
	\def\mn@doi{\begingroup\mn@urlcharsother \@ifnextchar [ {\mn@doi@}
		{\mn@doi@[]}}
	\def\mn@doi@[#1]#2{\def\@tempa{#1}\ifx\@tempa\@empty \href
		{http://dx.doi.org/#2} {doi:#2}\else \href {http://dx.doi.org/#2} {#1}\fi
		\endgroup}
	\def\mn@eprint#1#2{\mn@eprint@#1:#2::\@nil}
	\def\mn@eprint@arXiv#1{\href {http://arxiv.org/abs/#1} {{\tt arXiv:#1}}}
	\def\mn@eprint@dblp#1{\href {http://dblp.uni-trier.de/rec/bibtex/#1.xml}
		{dblp:#1}}
	\def\mn@eprint@#1:#2:#3:#4\@nil{\def\@tempa {#1}\def\@tempb {#2}\def\@tempc
		{#3}\ifx \@tempc \@empty \let \@tempc \@tempb \let \@tempb \@tempa \fi \ifx
		\@tempb \@empty \def\@tempb {arXiv}\fi \@ifundefined
		{mn@eprint@\@tempb}{\@tempb:\@tempc}{\expandafter \expandafter \csname
			mn@eprint@\@tempb\endcsname \expandafter{\@tempc}}}
	
	\bibitem[\protect\citeauthoryear{{Argall} et~al.,}{{Argall}
		et~al.}{2018}]{Argall2018ArXiv}
	{Argall} M.~R.,  et~al., 2018, arXiv e-prints, \href
	{https://ui.adsabs.harvard.edu/abs/2018arXiv180907388A} {p. arXiv:1809.07388}
	
	\bibitem[\protect\citeauthoryear{Bandyopadhyay et~al.,}{Bandyopadhyay
		et~al.}{2018}]{Bandyopadhyay2018bApJ}
	Bandyopadhyay R.,  et~al., 2018, \mn@doi [The Astrophysical Journal]
	{https://doi.org/10.3847/1538-4357/aade04}, 866, 106
	
	\bibitem[\protect\citeauthoryear{Banerjee \& Galtier}{Banerjee \&
		Galtier}{2013}]{Banerjee2013PRE}
	Banerjee S.,  Galtier S.,  2013, \mn@doi [Phys. Rev. E]
	{10.1103/PhysRevE.87.013019}, 87, 013019
	
	\bibitem[\protect\citeauthoryear{Batchelor}{Batchelor}{1953}]{Batchelor1953book}
	Batchelor G.~K.,  1953, The theory of homogeneous turbulence.
	Cambridge university press
	
	\bibitem[\protect\citeauthoryear{Biskamp}{Biskamp}{2003}]{Biskamp2003Book}
	Biskamp D.,  2003, Magnetohydrodynamic Turbulence.
	CUP, Cambridge, UK, \url
	{https://ui.adsabs.harvard.edu/abs/2008matu.book.....B}
	
	\bibitem[\protect\citeauthoryear{Borovsky \& Funsten}{Borovsky \&
		Funsten}{2003}]{Borovsky2003JGR}
	Borovsky J.~E.,  Funsten H.~O.,  2003, \mn@doi [Journal of Geophysical
	Research: Space Physics] {10.1029/2002JA009625}, 108
	
	\bibitem[\protect\citeauthoryear{Borovsky, Elphic, Funsten  \&
		Thomsen}{Borovsky et~al.}{1997}]{Borovsky1997JPP}
	Borovsky J.~E.,  Elphic R.~C.,  Funsten H.~O.,   Thomsen M.~F.,  1997, \mn@doi
	[Journal of Plasma Physics] {10.1017/S0022377896005259}, 57, 1
	
	\bibitem[\protect\citeauthoryear{{Bruno} \& {Carbone}}{{Bruno} \&
		{Carbone}}{2013}]{Bruno2013LRSP}
	{Bruno} R.,  {Carbone} V.,  2013, \mn@doi [Living Reviews in Solar Physics]
	{10.12942/lrsp-2013-2}, 10, 2
	
	\bibitem[\protect\citeauthoryear{Burch, Moore, Torbert  \& Giles}{Burch
		et~al.}{2016}]{Burch2016SSR}
	Burch J.~L.,  Moore T.~E.,  Torbert R.~B.,   Giles B.~L.,  2016, \mn@doi [Space
	Science Reviews] {10.1007/s11214-015-0164-9}, 199, 5
	
	\bibitem[\protect\citeauthoryear{Chasapis et~al.,}{Chasapis
		et~al.}{2017}]{Chasapis2017ApJ}
	Chasapis A.,  et~al., 2017, \mn@doi [Astrophys. J.]
	{10.3847/1538-4357/836/2/247}, 836, 247
	
	\bibitem[\protect\citeauthoryear{Chen, Leung, Boldyrev, Maruca  \& Bale}{Chen
		et~al.}{2014}]{ChenEA14}
	Chen C. H.~K.,  Leung L.,  Boldyrev S.,  Maruca B.~A.,   Bale S.~D.,  2014,
	\mn@doi [Geophys. Res. Lett.] {10.1002/2014GL062009}
	
	\bibitem[\protect\citeauthoryear{Dobrowolny, Mangeney  \& Veltri}{Dobrowolny
		et~al.}{1980}]{Dobrowolny1980PRL}
	Dobrowolny M.,  Mangeney A.,   Veltri P.,  1980, \mn@doi [Phys. Rev. Lett.]
	{10.1103/PhysRevLett.45.144}, 45, 144
	
	\bibitem[\protect\citeauthoryear{Eastwood, Phan, Bale  \& Tjulin}{Eastwood
		et~al.}{2009}]{EastwoodEA09}
	Eastwood J.~P.,  Phan T.~D.,  Bale S.~D.,   Tjulin A.,  2009, \mn@doi [Phys.
	Rev.~Lett.] {10.1103/PhysRevLett.102.035001}, 102
	
	\bibitem[\protect\citeauthoryear{Ergun et~al.,}{Ergun
		et~al.}{2018}]{Ergun2018GRL}
	Ergun R.~E.,  et~al., 2018, \mn@doi [Geophysical Research Letters]
	{10.1002/2018GL076993}, 45, 3338
	
	\bibitem[\protect\citeauthoryear{Frisch}{Frisch}{1995}]{Frisch1995Book}
	Frisch U.,  1995, Turbulence.
	Cambridge, UK
	
	\bibitem[\protect\citeauthoryear{Gershman et~al.,}{Gershman
		et~al.}{2018}]{Gershman2018PoP}
	Gershman D.~J.,  et~al., 2018, \mn@doi [Phys. Plasmas] {10.1063/1.5009158}, 25,
	022303
	
	\bibitem[\protect\citeauthoryear{Gershman et~al.,}{Gershman
		et~al.}{2019}]{Gershman2019JGR}
	Gershman D.~J.,  et~al., 2019, \mn@doi [Journal of Geophysical Research: Space
	Physics] {10.1029/2019JA026980}, 124, 10345
	
	\bibitem[\protect\citeauthoryear{Hadid, Sahraoui, Kiyani, Retinò, Modolo,
		Canu, Masters  \& Dougherty}{Hadid et~al.}{2015}]{Hadid2015ApJL}
	Hadid L.~Z.,  Sahraoui F.,  Kiyani K.~H.,  Retinò A.,  Modolo R.,  Canu P.,
	Masters A.,   Dougherty M.~K.,  2015, \mn@doi [Astrophys. J. Lett.]
	{10.1088/2041-8205/813/2/L29}, 813, L29
	
	\bibitem[\protect\citeauthoryear{Hadid, Sahraoui, Galtier  \& Huang}{Hadid
		et~al.}{2018}]{Hadid2018PRL}
	Hadid L.~Z.,  Sahraoui F.,  Galtier S.,   Huang S.~Y.,  2018, \mn@doi [Phys.
	Rev. Lett.] {10.1103/PhysRevLett.120.055102}, 120, 055102
	
	\bibitem[\protect\citeauthoryear{Huang et~al.,}{Huang
		et~al.}{2012}]{Huang2012GRL}
	Huang S.~Y.,  et~al., 2012, \mn@doi [Geophysical Research Letters]
	{10.1029/2012GL052210}, 39, L11104
	
	\bibitem[\protect\citeauthoryear{Huang, Sahraoui, Deng, He, Yuan, Zhou, Pang
		\& Fu}{Huang et~al.}{2014}]{Huang2014ApJL}
	Huang S.~Y.,  Sahraoui F.,  Deng X.~H.,  He J.~S.,  Yuan Z.~G.,  Zhou M.,  Pang
	Y.,   Fu H.~S.,  2014, \mn@doi [The Astrophysical Journal Letters]
	{10.1088/2041-8205/789/2/l28}, 789, L28
	
	\bibitem[\protect\citeauthoryear{Kolmogorov}{Kolmogorov}{1941a}]{Kolmogorov1941a}
	Kolmogorov A.~N.,  1941a, \mn@doi [Dokl. Akad. Nauk SSSR]
	{10.1098/rspa.1991.0075}, 30, 301
	
	\bibitem[\protect\citeauthoryear{Kolmogorov}{Kolmogorov}{1941b}]{Kolmogorov1941c}
	Kolmogorov A.~N.,  1941b, \mn@doi [C.R. Acad. Sci. U.R.S.S.]
	{10.1098/rspa.1991.0076}, 32, 16
	
	\bibitem[\protect\citeauthoryear{{Le Contel} et~al.,}{{Le Contel}
		et~al.}{2016}]{LeContel2016SSR}
	{Le Contel} O.,  et~al., 2016, \mn@doi [Space Science Reviews]
	{10.1007/s11214-014-0096-9}, \href
	{https://ui.adsabs.harvard.edu/abs/2016SSRv..199..257L} {199, 257}
	
	\bibitem[\protect\citeauthoryear{MacBride, Smith  \& Forman}{MacBride
		et~al.}{2008}]{MacBride2008ApJ}
	MacBride B.~T.,  Smith C.~W.,   Forman M.~A.,  2008, \mn@doi [Astrophys. J.]
	{10.1086/529575}, 679, 1644
	
	\bibitem[\protect\citeauthoryear{Matthaeus \& Goldstein}{Matthaeus \&
		Goldstein}{1982}]{Matthaeus1982aJGR}
	Matthaeus W.~H.,  Goldstein M.~L.,  1982, \mn@doi [J. Geophys. Res.]
	{10.1029/JA087iA08p06011}, 87, 6011
	
	\bibitem[\protect\citeauthoryear{{Matthaeus} \& {Lamkin}}{{Matthaeus} \&
		{Lamkin}}{1986}]{Matthaeus1986PoF}
	{Matthaeus} W.~H.,  {Lamkin} S.~L.,  1986, \mn@doi [Physics of Fluids]
	{10.1063/1.866004}, \href
	{https://ui.adsabs.harvard.edu/abs/1986PhFl...29.2513M} {29, 2513}
	
	\bibitem[\protect\citeauthoryear{{Matthaeus} \& {Velli}}{{Matthaeus} \&
		{Velli}}{2011}]{Matthaeus2011SSR}
	{Matthaeus} W.~H.,  {Velli} M.,  2011, \mn@doi [Space Science Reviews]
	{10.1007/s11214-011-9793-9}, 160, 145
	
	\bibitem[\protect\citeauthoryear{Monin \& Yaglom}{Monin \&
		Yaglom}{1971}]{MoninYaglom-vol1}
	Monin A.~S.,  Yaglom A.~M.,  1971, Statistical Fluid Mechanics, Vol.~1.
	MIT Press, Cambridge, Mass.
	
	\bibitem[\protect\citeauthoryear{Monin \& Yaglom}{Monin \&
		Yaglom}{1975}]{MoninYaglom-vol2}
	Monin A.~S.,  Yaglom A.~M.,  1975, Statistical Fluid Mechanics, Vol.~2.
	MIT Press, Cambridge, Mass., \url
	{https://ui.adsabs.harvard.edu/abs/1971sfmm.book.....M}
	
	\bibitem[\protect\citeauthoryear{Osman, Wan, Matthaeus, Weygand  \&
		Dasso}{Osman et~al.}{2011}]{Osman2011PRL}
	Osman K.~T.,  Wan M.,  Matthaeus W.~H.,  Weygand J.~M.,   Dasso S.,  2011,
	\mn@doi [Phys. Rev. Lett.] {10.1103/PhysRevLett.107.165001}, 107, 165001
	
	\bibitem[\protect\citeauthoryear{Osman, Kiyani, Matthaeus, Hnat, Chapman  \&
		Khotyaintsev}{Osman et~al.}{2015}]{Osman2015ApJL}
	Osman K.~T.,  Kiyani K.~H.,  Matthaeus W.~H.,  Hnat B.,  Chapman S.~C.,
	Khotyaintsev Y.~V.,  2015, \mn@doi [The Astrophysical Journal Letters]
	{https://doi.org/10.1088/2041-8205/815/2/L24}, 815, L24
	
	\bibitem[\protect\citeauthoryear{Oughton, Priest  \& Matthaeus}{Oughton
		et~al.}{1994}]{Oughton1994JFM}
	Oughton S.,  Priest E.~R.,   Matthaeus W.~H.,  1994, \mn@doi [J. Fluid Mech.]
	{10.1017/S0022112094002867}, 280, 95
	
	\bibitem[\protect\citeauthoryear{Parker}{Parker}{1957}]{Parker1957JGR}
	Parker E.~N.,  1957, \mn@doi [Journal of Geophysical Research (1896-1977)]
	{10.1029/JZ062i004p00509}, 62, 509
	
	\bibitem[\protect\citeauthoryear{{Parker}}{{Parker}}{1979}]{Parker1979book}
	{Parker} E.~N.,  1979, {Cosmical magnetic fields: Their origin and their
		activity}.
	PB - Oxford University Press
	
	\bibitem[\protect\citeauthoryear{Politano \& Pouquet}{Politano \&
		Pouquet}{1998a}]{Politano1998GRL}
	Politano H.,  Pouquet A.,  1998a, \mn@doi [Geophysical Research Letters]
	{10.1029/97GL03642}, 25, 273
	
	\bibitem[\protect\citeauthoryear{Politano \& Pouquet}{Politano \&
		Pouquet}{1998b}]{Politano1998PRE}
	Politano H.,  Pouquet A.,  1998b, \mn@doi [Phys. Rev. E]
	{10.1103/PhysRevE.57.R21}, 57, R21
	
	\bibitem[\protect\citeauthoryear{Pollock et~al.,}{Pollock
		et~al.}{2016}]{Pollock2016SSR}
	Pollock C.,  et~al., 2016, \mn@doi [Space Science Reviews]
	{10.1007/s11214-016-0245-4}, 199, 331
	
	\bibitem[\protect\citeauthoryear{Pucci et~al.,}{Pucci
		et~al.}{2017}]{Pucci2017ApJ}
	Pucci F.,  et~al., 2017, \mn@doi [The Astrophysical Journal]
	{https://doi.org/10.3847/1538-4357/aa704f}, 841, 60
	
	\bibitem[\protect\citeauthoryear{Richardson}{Richardson}{1922}]{Richardson1922book}
	Richardson L.~F.,  1922, Weather prediction by numerical process.
	Cambridge university press, \url
	{https://books.google.com/books?hl=en&lr=&id=D52d3_bbgg8C&oi=fnd&pg=PA3&ots=60SBkUYITo&sig=Gd0SoSuVZ7ndNA-o2c14DmPdvKE}
	
	\bibitem[\protect\citeauthoryear{Russell et~al.,}{Russell
		et~al.}{2016}]{Russell2016SSR}
	Russell C.~T.,  et~al., 2016, \mn@doi [Space Science Reviews]
	{10.1007/s11214-014-0057-3}, 199, 189
	
	\bibitem[\protect\citeauthoryear{{Saur}}{{Saur}}{2004}]{Saur2004ApJL}
	{Saur} J.,  2004, \mn@doi [The Astrophysical Journal Letters] {10.1086/382588},
	\href {https://ui.adsabs.harvard.edu/abs/2004ApJ...602L.137S} {602, L137}
	
	\bibitem[\protect\citeauthoryear{Schindler, Hesse  \& Birn}{Schindler
		et~al.}{1988}]{SchindlerEA88}
	Schindler K.,  Hesse M.,   Birn J.,  1988, \mn@doi [J.~Geophys.~Res.]
	{10.1029/JA093iA06p05547}, 93, 5547
	
	\bibitem[\protect\citeauthoryear{Servidio, Matthaeus, Shay, Cassak  \&
		Dmitruk}{Servidio et~al.}{2009}]{Servidio2009PRL}
	Servidio S.,  Matthaeus W.~H.,  Shay M.~A.,  Cassak P.~A.,   Dmitruk P.,  2009,
	\mn@doi [Phys. Rev. Lett.] {10.1103/PhysRevLett.102.115003}, 102, 115003
	
	\bibitem[\protect\citeauthoryear{Shay, Haggerty, Matthaeus, Parashar, Wan  \&
		Wu}{Shay et~al.}{2018}]{Shay2018PoP}
	Shay M.~A.,  Haggerty C.~C.,  Matthaeus W.~H.,  Parashar T.~N.,  Wan M.,   Wu
	P.,  2018, \mn@doi [Physics of Plasmas] {10.1063/1.4993423}, 25, 012304
	
	\bibitem[\protect\citeauthoryear{Shebalin, Matthaeus  \& Montgomery}{Shebalin
		et~al.}{1983}]{Shebalin1983JPP}
	Shebalin J.~V.,  Matthaeus W.~H.,   Montgomery D.,  1983, \mn@doi [J. Plasma
	Phys.] {10.1017/S0022377800000933}, 29, 525
	
	\bibitem[\protect\citeauthoryear{Sonnerup}{Sonnerup}{1984}]{SonnerupEA84}
	Sonnerup B. U. O. e.~a.,  1984. NASA Reference Publication 1120
	
	\bibitem[\protect\citeauthoryear{Sorriso-Valvo et~al.,}{Sorriso-Valvo
		et~al.}{2007}]{Sorriso-Valvo2007PRL}
	Sorriso-Valvo L.,  et~al., 2007, \mn@doi [Phys. Rev. Lett.]
	{10.1103/PhysRevLett.99.115001}, 99, 115001
	
	\bibitem[\protect\citeauthoryear{Sorriso-Valvo et~al.,}{Sorriso-Valvo
		et~al.}{2019}]{Sorriso-Valvo2019PRL}
	Sorriso-Valvo L.,  et~al., 2019, \mn@doi [Phys. Rev. Lett.]
	{10.1103/PhysRevLett.122.035102}, 122, 035102
	
	\bibitem[\protect\citeauthoryear{Stawarz, Smith, Vasquez, Forman  \&
		Mac{Bride}}{Stawarz et~al.}{2009}]{Stawarz2009ApJ}
	Stawarz J.~E.,  Smith C.~W.,  Vasquez B.~J.,  Forman M.~A.,   Mac{Bride} B.~T.,
	2009, \mn@doi [ApJ] {10.1088/0004-637X/697/2/1119}, 697, 1119
	
	\bibitem[\protect\citeauthoryear{Stawarz et~al.,}{Stawarz
		et~al.}{2016}]{Stawarz2016JGR}
	Stawarz J.~E.,  et~al., 2016, \mn@doi [Journal of Geophysical Research: Space
	Physics] {10.1002/2016JA023458}, 121, 11,021
	
	\bibitem[\protect\citeauthoryear{{Stepanova}, {Antonova}, {Paredes-Davis},
		{Ovchinnikov}  \& {Yermolaev}}{{Stepanova} et~al.}{2009}]{Stepanova2009AG}
	{Stepanova} M.,  {Antonova} E.~E.,  {Paredes-Davis} D.,  {Ovchinnikov} I.~L.,
	{Yermolaev} Y.~I.,  2009, \mn@doi [Annales Geophysicae]
	{10.5194/angeo-27-1407-2009}, \href
	{https://ui.adsabs.harvard.edu/abs/2009AnGeo..27.1407S} {27, 1407}
	
	\bibitem[\protect\citeauthoryear{Taylor}{Taylor}{1938}]{Taylor1938PRSLA}
	Taylor G.~I.,  1938, \mn@doi [Proceedings of the Royal Society of London Series
	A] {10.1098/rspa.1938.0032}, 164, 476
	
	\bibitem[\protect\citeauthoryear{{Taylor}}{{Taylor}}{1986}]{Taylor86}
	{Taylor} J.~B.,  1986, \mn@doi [Reviews of Modern Physics]
	{10.1103/RevModPhys.58.741}, \href
	{http://adsabs.harvard.edu/abs/1986RvMP...58..741T} {58, 741}
	
	\bibitem[\protect\citeauthoryear{{Vasyliunas}}{{Vasyliunas}}{1975}]{Vasyliunas1975JGR}
	{Vasyliunas} V.~M.,  1975, \mn@doi [Reviews of Geophysics and Space Physics]
	{10.1029/RG013i001p00303}, \href
	{https://ui.adsabs.harvard.edu/abs/1975RvGSP..13..303V} {13, 303}
	
	\bibitem[\protect\citeauthoryear{Verma}{Verma}{1996}]{Verma1996JGR}
	Verma M.~K.,  1996, \mn@doi [Journal of Geophysical Research: Space Physics]
	{10.1029/96JA02324}, 101, 27543
	
	\bibitem[\protect\citeauthoryear{Verma, Roberts  \& Goldstein}{Verma
		et~al.}{1995}]{Verma1995JGR}
	Verma M.~K.,  Roberts D.~A.,   Goldstein M.~L.,  1995, \mn@doi [Journal of
	Geophysical Research: Space Physics] {10.1029/95JA01216}, 100, 19839
	
	\bibitem[\protect\citeauthoryear{{Verscharen}, {Klein}  \&
		{Maruca}}{{Verscharen} et~al.}{2019}]{Verscharen2019LRSP}
	{Verscharen} D.,  {Klein} K.~G.,   {Maruca} B.~A.,  2019, \mn@doi [Living
	Reviews in Solar Physics] {10.1007/s41116-019-0021-0}, 16, 5
	
	\bibitem[\protect\citeauthoryear{V{\"o}r{\"o}s et~al.,}{V{\"o}r{\"o}s
		et~al.}{2004}]{Voros2004JGR}
	V{\"o}r{\"o}s Z.,  et~al., 2004, \mn@doi [Journal of Geophysical Research:
	Space Physics] {10.1029/2004JA010404}, 109
	
	\bibitem[\protect\citeauthoryear{Zimbardo, Greco, Sorriso-Valvo, Perri,
		V{\"o}r{\"o}s, Aburjania, Chargazia  \& Alexandrova}{Zimbardo
		et~al.}{2010}]{Zimbardo2010SSR}
	Zimbardo G.,  Greco A.,  Sorriso-Valvo L.,  Perri S.,  V{\"o}r{\"o}s Z.,
	Aburjania G.,  Chargazia K.,   Alexandrova O.,  2010, \mn@doi [Space Science
	Reviews] {10.1007/s11214-010-9692-5}, 156, 89
	
	\bibitem[\protect\citeauthoryear{de K{\'a}rm{\'a}n \&
		Howarth}{de~K{\'a}rm{\'a}n \& Howarth}{1938}]{Karman1938PRSL}
	de K{\'a}rm{\'a}n T.,  Howarth L.,  1938, \mn@doi [Proc. Roy. Soc. London Ser.
	A] {10.1098/rspa.1938.0013}, 164, 192
	
	\makeatother
\end{thebibliography}

%\begin{thebibliography}{99}
%\end{thebibliography}

%%%%%%%%%%%%%%%%%%%%%%%%%%%%%%%%%%%%%%%%%%%%%%%%%%

%%%%%%%%%%%%%%%%% APPENDICES %%%%%%%%%%%%%%%%%%%%%

%\appendix


%If you want to present additional material which would interrupt the flow of the main paper, it can be placed in an Appendix which appears after the list of references.

%%%%%%%%%%%%%%%%%%%%%%%%%%%%%%%%%%%%%%%%%%%%%%%%%%

% Don't change these lines
%\bsp	% typesetting comment
\label{lastpage}
\end{document}